\documentclass[aps,prc,reprint,superscriptaddress,nofootinbib,floatfix,showkeys]{revtex4-2}

\usepackage[utf8]{inputenc}
\usepackage[T1]{fontenc}
\usepackage{xfrac}
\usepackage{amsmath}
\usepackage{amssymb}
\usepackage{braket}
\usepackage{graphicx}
\usepackage{xcolor}

\usepackage{lineno}
\modulolinenumbers[5]

\usepackage{hyperref}
\hypersetup{colorlinks=true,
  citecolor=blue,linkcolor=blue}
\usepackage[capitalise]{cleveref}

\newcommand{\ba}{\begin{align}}
\newcommand{\ea}{\end{align}}

\begin{document}

\title{Intertwined quantum phase transitions in the even-even $^{90\text{--}100}$Sr isotopes}

\author{N.~Gavrielov}
\email{noamgavrielov@gmail.com}

\affiliation{School of Physics and Astronomy, Tel Aviv 
University, Tel Aviv, Israel}
\affiliation{Department of Physics, Nuclear Research Center 
Negev, Be'er Sheva 84190, Israel}

\date{\today}

\begin{abstract}
The even-even $^{90\text{--}100}$Sr isotopes are identified 
as a region of intertwined quantum phase transitions 
(IQPTs). In this scenario, a quantum phase transition 
involving the crossing of normal and intruder 
configurations is accompanied by a shape evolution within 
the intruder configuration. 
Using the interacting boson model with configuration mixing 
(IBM-CM), its is shown that the strontium chain exhibits 
the IQPT scenario, where the intruder configuration evolves 
from a near-spherical structure in $^{90\text{--}96}$Sr to 
a deformed one in $^{98,100}$Sr, while the normal and intruder 
configurations cross between $^{96}$Sr and $^{98}$Sr. As a 
result, the ground state changes abruptly from a weakly 
collective normal configuration to a deformed intruder 
configuration. Evidence for this scenario is provided by a detailed 
comparison with experimental excitation energies, 
isotope shifts, and monopole $E0$ transition strengths, 
together with the configuration and $n_d$ decompositions 
of the calculated wave functions. The results place the strontium  
isotopes alongside the neighboring zirconium chain as 
a realization of IQPTs in the intricate $A\approx100$ 
region.
\end{abstract}


\maketitle

%\linenumbers

Nuclei in the region of $A \approx 100$ are known to 
exhibit abrupt structural changes around neutron number $N 
= 60$. For neutron numbers less than 60, the ground state 
configuration was traditionally interpreted as excitations 
of a spherical vibrator, arising from different 
single-particle excitations, either proton or neutron 
\cite{Federman1977, Federman1979, Leoni2024}. For neutron 
number 60 and larger, the ground state configuration 
exhibits dense spectra, 
associated with deformed rotors of either prolate or oblate 
shapes, which are manifested due to the increase of 
collectivity. This onset of deformation near $N\!=\!60$ is 
among the most rapid in the nuclear chart.
However, such evolution in shape is also associated 
with a crossing between different proton configurations 
\cite{Federman1979}, leading us to identify 
different coexisting shapes in the low-lying spectra 
\cite{Leoni2024, Garrett2022}.
These structural changes are also identified as quantum 
phase transitions (QPTs) \cite{Cejnar2010} in different 
isotopes in this region, with neutron number 60 being the 
epicenter of the abrupt structural changes.

QPTs in nuclei involve abrupt structural changes driven by
the variation of a discrete control parameter, the nucleon 
number, at zero temperature. In algebraic descriptions, 
they arise from changes in the topology of the underlying
energy surface and in the character of relevant eigenstates
\cite{Gilmore1978a, Gilmore1979}.
However, these changes can occur due to two main 
mechanisms, a shape evolution within a single shell-model 
configuration, also known as Type~I QPT, and a crossing of 
multiple shell-model configurations, also known as Type~II 
QPT.

A Type~I QPT occurs when one modifies the number of 
nucleons, e.g. increasing the number of neutrons, and the 
shape of the isotopes evolves from one to another, e.g. 
spherical to deformed. 
A Type~II~QPT \cite{Frank2006} occurs when 
protons and neutrons that occupy spin-orbit partner 
orbitals, 
$\pi(n\ell_{\ell\pm1/2})$--$\nu(n\ell_{\ell\mp1/2})$,
interact via the residual isoscalar proton-neutron 
interaction, $V_{pn}$~\cite{Heyde2011,Federman1979}. 
The resulting gain in $n$-$p$ energy compensates the loss 
in single-particle and pairing energy. As a consequence, a 
mutual polarization effect occurs, which lowers 
single-particle orbitals of higher configurations to near 
(and effectively below) the ground state configuration. 
If the mixing is small, the Type~II~QPT can be accompanied 
by a distinguished Type~I~QPT within each configuration 
separately. Such a scenario is referred to as intertwined 
quantum phase transitions (IQPTs). It was recently shown to 
occur in the even-even zirconium isotopes 
\cite{Gavrielov2019, Gavrielov2020, Gavrielov2022}, where 
the normal configuration remains spherical while the 
intruder undergoes its own shape evolution with increasing 
neutron number, accompanied by a crossing between the two 
configurations.
The occurrence of IQPTs was subsequently demonstrated in 
odd-mass niobium \cite{Gavrielov2022c, Gavrielov2023a} and 
zirconium isotopes \cite{Gavrielov2025}, underscoring the 
necessity of incorporating multiple configurations and 
analyzing their individual shape evolution to understand 
the structure of nuclei in the $A\!\sim\!100$ region.

The strontium isotopes with $Z\!=\!38$ lie two protons 
below the $Z\!=\!40$ subshell closure and share many of 
the structural characteristics of the zirconium chain for 
neutron number $N=52\text{--}62$, where experimental data 
are available. 
Experimental studies of $^{90\text{--}96}$Sr have 
established weak $B(E2;2^+_1\to0^+_1)$ values 
\cite{Mach1991}, suggesting a single-particle 
character, while Coulomb excitation measurements of
$^{96,98}$Sr have suggested the coexistence of a deformed 
prolate configuration with a spherical one near $N\!=\!60$, 
with very small mixing between them 
\cite{Clement2016,Clement2016a}.
Theoretical descriptions of the strontium isotopes around 
$A \approx 100$ include IBM-CM calculations 
\cite{MayaBarbecho2022}, mapped IBM studies based on the
Gogny energy density functional \cite{Nomura2016c}, and 
Monte Carlo shell-model (MCSM) calculations 
\cite{Regis2017c}.
These works established the abrupt structural change and 
shape coexistence near $N=60$ from complementary 
microscopic and collective perspectives. 

The present analysis follows the approach developed for the 
even-even zirconium isotopes \cite{Gavrielov2019, 
Gavrielov2020, Gavrielov2022} and focuses on the  
transparent identification of IQPTs in the strontium chain, 
thereby placing them alongside the zirconium chain in a 
unified picture of this region.

The $^{90\text{--}100}$Sr isotopes are described in the 
IBM-CM with two configurations. The IBM describes low-lying
quadrupole states in even-even nuclei in terms of
a system of monopole ($s$) and quadrupole ($d$) bosons 
representing valence nucleon pairs. The first 
configuration, normal ($A$), corresponds to $0p$-$2h$ 
proton excitations with respect to the $Z\!=\!40$ subshell, 
giving a boson number of $N_A \!\equiv\! N_\pi\! + 
N_\nu\!=\! N_b$~.
The second configuration, intruder ($B$), corresponds to 
$2p$-$4h$ proton excitations across the same subshell, 
giving $N_B \! \equiv \! (N_\pi\!+\!2) \!+\! N_\nu \!=\! 
N_b\!+\!2$ bosons.

The total Hamiltonian takes the block form \cite{Duval1981, 
Duval1982}
\begin{equation}
\hat H = \begin{bmatrix}
  \hat H^{(A)}(\xi^{(A)}) &
  \hat W(\omega) \\
  \hat W(\omega) &
  \hat H^{(B)}(\xi^{(B)})
\end{bmatrix}~,
\label{eq:H}
\end{equation}
where $\hat H^{(A)}$ and $\hat H^{(B)}$
act on the $[N_A\!=\!N_b]$ and $[N_B\!=\!N_b+2]$ 
boson spaces, respectively, and $\hat W$ is the mixing 
interaction
\begin{equation}\label{eq:W}
\hat W = \omega\bigl[(d^\dagger \times d^\dagger)^{(0)}
  + (s^\dagger)^2\bigr] + \text{H.c.}~,
\end{equation}
which connects states of the same angular momentum from the 
two boson spaces.
The configuration Hamiltonians are
\begin{subequations}
\label{eq:HAB}
\begin{align}
\hat H^{(A)} &=
  \epsilon_d^{(A)}\hat n_d
  + \kappa^{(A)}\hat Q_\chi\cdot
  \hat Q_\chi
  + \kappa^{\prime(A)}
  \hat L\cdot\hat L~, \label{eq:HA}\\
\hat H^{(B)} &=
  \epsilon_d^{(B)}\hat n_d
  + \kappa^{(B)}\hat Q_\chi\cdot
  \hat Q_\chi
  + \kappa^{\prime(B)}
  \hat L\cdot\hat L
  + \Delta_p~, \label{eq:HB}
\end{align}
\end{subequations}
where $\hat n_d$ is the $d$-boson number
operator, $\hat Q_\chi \!=\! (s^\dagger
\tilde d\!+\!d^\dagger s)^{(2)}\!+\!
\chi(d^\dagger\tilde d)^{(2)}$, $\hat L$
is the angular momentum operator, and
$\Delta_p$ is the energy offset of the
intruder configuration, reflecting the
monopole contribution of the proton-neutron
interaction \cite{Heyde1985,Heyde1987}.

After diagonalization of
Eq.~(\ref{eq:H}), the eigenstates
$\ket{\Psi;L}$ are expanded as
\begin{multline}\label{eq:wf}
\ket{\Psi;L} = \sum_{\alpha}  C^{(N_b,L)}_{\alpha}
  \ket{\Psi_A;[N_b],\alpha,L} \\
  + \sum_{\alpha} C^{(N_b+2,L)}_{\alpha}
  \ket{\Psi_B;[N_b+2],\alpha,L}~,
\end{multline}
where $\alpha$ denotes additional quantum numbers 
characterizing the boson basis, and $L$ is the total
angular momentum. The occupation probabilities for the 
boson number $N_b$ and the $d$-boson number $n_d$  
of the wave function can be determined from
\begin{subequations}
\begin{align}
v^2_{(N_i,L)} &=
  \sum_{n_d} v^2_{(n_d;N_i,L)},
  \quad i = A, B~,
\label{eq:prob_bos}\\
v^2_{(n_d;N_i,L)} &=
  \sum_{\tau,n_\Delta}
  \left|C^{(N_i,L)}_{n_d,\tau,
  n_\Delta}\right|^2~,
\label{eq:prob_nd}
\end{align}
\end{subequations}
where $N_A\!=\!N_b$, $N_B\!=\!N_b\!+\!2$ and 
$v^2_{(N_A,L)}+v^2_{(N_B,L)}\!=\!1$.
\Cref{eq:prob_bos} gives the occupation probabilities of the normal 
($v^2_{(N_A,L)}$) and intruder ($v^2_{(N_B,L)}$) 
configurations in the wave function. \Cref{eq:prob_nd} gives 
the occupation probability of the number of $d$-bosons 
$n_d$, which indicates the degree of deformation.
A dominant single $n_d$ component corresponds to a U(5) 
dynamical symmetry (DS) wave function, equivalent to a 
spherical shape (phase) in the geometrical 
interpretation of the IBM 
\cite{Ginocchio1980a, Ginocchio1980b, Dieperink1980}.
A spread over multiple $n_d$ values indicates
deformation.

\begin{table}[t!]
\caption{Parameters of the IBM-CM
Hamiltonian~(\ref{eq:HAB})
in MeV; $\chi^{(A,B)}$ are dimensionless
and $e_B^{(A,B)}$ are in
$\sqrt{\text{W.u.}}$.
The first row lists the neutron number
and normal-intruder boson numbers $(N,N+2)$.
\label{tab:parameters}}
\centering
\begingroup
\footnotesize
\setlength{\tabcolsep}{1.8pt}
\renewcommand{\arraystretch}{1.0}
\begin{tabular}{@{}lcccccc@{}}
\hline\hline
  & $52\,(2,4)$
  & $54\,(3,5)$
  & $56\,(4,6)$
  & $58\,(5,7)$
  & $60\,(6,8)$
  & $62\,(7,9)$\\
\hline
$\epsilon_d^{(A)}$
  & 0.79 & 0.79 & 0.79
  & 0.79 & 0.79 & 0.79 \\
$\kappa^{(A)}$
  & $-$0.00238 & $-$0.00238
  & $-$0.00238 & $-$0.00238
  & $-$0.00238 & $-$0.00238 \\
$\kappa^{\prime(A)}$
  & 0.01 & 0.01 & 0.01
  & 0.01 & 0.01 & 0.01 \\
$\chi^{(A)}$
  & $+1$ & $+1$ & $+1$
  & $+1$ & $+1$ & $+1$ \\
$\epsilon_d^{(B)}$
  & 0.35 & 0.35 & 0.35
  & 0.35 & 0.25 & 0.15 \\
$\kappa^{(B)}$
  & $-$0.015 & $-$0.015
  & $-$0.015 & $-$0.015
  & $-$0.015 & $-$0.015 \\
$\kappa^{\prime(B)}$
  & 0.03 & 0.03 & 0.01
  & 0.01 & 0.01 & 0.01 \\
$\chi^{(B)}$
  & $+1$ & $+1$ & $+1$
  & $-1$ & $-1$ & $-1$ \\
$\Delta_p$
  & 3.0 & 2.6 & 2.4
  & 1.8 & 0.8 & 0.8 \\
$\omega$
  & 0.017 & 0.017 & 0.017
  & 0.017 & 0.017 & 0.017 \\
$e_B^{(A)}$
  & 2.050 & 1.625 & 1.400
  & 1.555 & 1.870 & 1.870 \\
$e_B^{(B)}$
  & 2.240 & 2.240 & 2.240
  & 2.240 & 2.240 & 1.825 \\
\hline\hline
\end{tabular}
\endgroup
\end{table}

\begin{figure*}[tb!]
\centering
\includegraphics[width=1\linewidth]{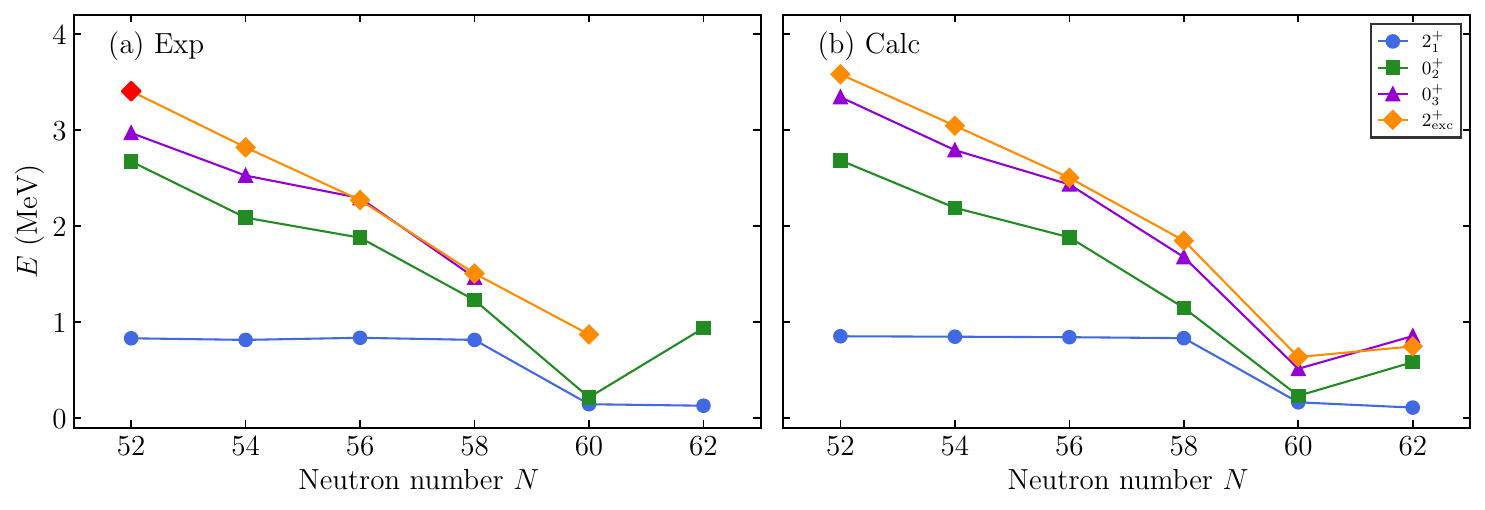}
\caption{Comparison between experimental and calculated 
lowest-energy levels in even-even strontium isotopes. For 
$^{90}$Sr, the red $2^+_\text{exc}$ marker denotes an 
extrapolated experimental value. Data taken from 
\cite{ensdf,Cruz2020} for $^{94}$Sr.
\label{fig:energies}}
\end{figure*}

The parameters used in this work are listed in 
Table~\ref{tab:parameters}. The boson numbers are 
indicated as $(N_b,N_b+2)$ with $N_b\!=\!N_\pi+N_\nu$, 
where $N_\pi\!=\!1$ and $N_\nu\!=\!1$--$6$ for 
$^\text{90--100}$Sr. The normal configuration parameters 
are held constant across the chain, consistent with the 
stability of its weakly collective character.
The intruder configuration is governed by a much stronger 
quadrupole interaction, with
\mbox{$\kappa^{(A)} \!\approx\! \sfrac{1}{6} \kappa^{(B)}$.}
This ratio is in the same spirit as the empirical 
relationship
\mbox{$\kappa^{(A)}\!\approx\!\sfrac{1}{3}\kappa^{(B)}$} 
observed in the zirconium chain 
\cite{Gavrielov2019, Gavrielov2022},
indicating in both cases the stronger collective tendency 
of the intruder configuration. The absolute value of the 
parameter $\chi^{(A,B)}$ is taken constant throughout the 
chain, $|\chi^{(A,B)}| \!=\! 1$. Its sign is determined 
using the results of Ref.~\cite{Marchini2026}, where the 
spectroscopic quadrupole moments of $^{94}$Zr were 
measured, suggesting an oblate shape for the intruder 
states. Therefore we take a positive sign for 
neutron number 52--56 ($^{90\text{--}94}$Sr), where the 
subshell $\nu d_{5/2}$ is being filled in a naive shell 
model counting scheme, and is rather separated from the 
$\nu s_{1/2}$, $\nu d_{3/2}$, $\nu g_{7/2}$, $\nu h_{11/2}$ 
orbits lying above it. For $^{96\text{--}98}$Sr, 
Refs.~\cite{Clement2016, Clement2016a} suggest a prolate 
deformation, which sets the intruder $\chi^{(B)}$ to be 
negative, and it is kept that way up to $^{100}$Sr.
The $\epsilon_d^{(B)}$ parameter, responsible for the 
spherical part of the intruder configuration, is constant 
at $0.35$~MeV for neutron numbers 52--58 and decreases to 
0.25 and 0.15~MeV at 60--62, respectively, due to the onset 
of deformation within the intruder configuration.
The energy offset $\Delta_p$ decreases gradually 
from $3.0$~MeV at neutron number 52 to 1.8~MeV at 58. Then, 
at 60--62, it undergoes a sudden drop to $0.8$~MeV due to 
the Type~II QPT. This reduction encodes the progressive 
lowering of the intruder configuration by the 
proton-neutron monopole interaction 
\cite{Heyde1985,Heyde1987}, and follows almost 
exactly the same values as the even-even zirconium 
calculation \cite{Gavrielov2019,Gavrielov2022}, suggesting 
the microscopic interpretation is consistent in this region.
The $\kappa^{\prime(B)}$ term is larger ($0.03$~MeV) for 
$N\!=\!52$--$54$ and is $0.01$~MeV from $N\!=\!56$
onward.
The mixing parameter $\omega\!=\!0.017$~MeV remains 
constant, and its value is consistent with that of the 
zirconium chain (0.02~MeV) 
\cite{Gavrielov2019, Gavrielov2020, Gavrielov2022}.
Overall, we see a very simple and transparent behavior 
of the parameters in which most of them are kept constant 
or have a well-defined trend.

The $E2$ transition operator used to calculate the rates is
\begin{equation}\label{eq:TE2}
\hat T(E2) \!=\! 
e_B^{(A)} \hat Q^{(N_b)}_\chi \!+\! e_B^{(B)} \hat 
Q^{(N_b+2)}_\chi~,
\end{equation}
where the superscript denotes a projection onto the corresponding boson space.
The effective charges $e_B^{(A,B)}$ were determined from the lowest
$2^+ \to 0^+$ transitions within the normal and intruder configurations.  

In order to illuminate the occurrence of IQPTs in the 
strontium isotopes in the framework of the IBM-CM, it is 
paramount to examine the behavior of multiple experimental 
observables and compare them to the calculation. This will 
also help validate the choice of parameters. In this 
work we examine the evolution of energy levels, $E2$ transition 
rates, isotope shifts and monopole strength of $E0$ transitions.

The calculated energy levels are obtained by diagonalizing 
the IBM-CM Hamiltonian. Their evolution traces both the 
relative motion of the normal and intruder configurations 
and the development of collectivity within each 
configuration. 

Figure~\ref{fig:energies} shows the evolution of selected 
experimental and calculated levels across the chain.
For $^{90}$Sr, the red marker indicates an extrapolated 
location of the intruder $2^+$ band member, inferred from 
the systematic evolution of the intruder configuration, and 
is shown only as a guide to the eye.

For $^{90,92}$Sr, we first note a caveat. The three 
low-lying $2^+$ states above the $2^+_1$ level and below 
the intruder $0^+_2$ state are not shown in the comparison. 
These states are not assigned here to the collective normal 
or intruder configurations of the present IBM-CM model 
space. This choice is motivated by the interpretation of 
Federman and collaborators~\cite{Federman1984} for
the low-lying $2^+_1$ state in terms of proton $1p$-$1h$ 
excitations from the $\pi p_{3/2}$ to the $\pi p_{1/2}$ 
orbit, and by QRPA calculations \cite{Severyukhin2018}
in which the $2^+_2$ state of $^{90}$Sr was identified as a 
proton-neutron mixed-symmetry excitation with 
configurational isospin polarization. 
%%%%%%%%%%%%%%%%%%%%%%%%%%%%%%%%%%%%%%%%%%%%%%%%%%%%%%%%%%%%%%%%%%%
\iffalse
Furthermore, in a naive shell-model picture, the $\pi (p_{3/2}^{-1} 
p_{1/2}^1)_{J_\pi = 1,2}$ proton excitations can be coupled 
to the neutron configuration $\nu (d_{5/2})^{n}_{J_\nu = 
0,2,4}$, with $n=2,4$ for $^{90,92}$Sr. This gives four 
possible $J=2^+$ states in this restricted schematic space. 
\fi
%%%%%%%%%%%%%%%%%%%%%%%%%%%%%%%%%%%%%%%%%%%%%%%%%%%%%%%%%%%%%%%%%%%
We therefore treat the remaining low-lying $2^+$ states as 
likely belonging to the same class of excitations. Such 
components require explicit proton-neutron or shell-model 
degrees of freedom, which are not included in the present 
two-configuration IBM-CM Hamiltonian.

For the IBM-CM calculation, the $2^+_1$ energy remains 
nearly constant at $\approx\!0.83$~MeV for $N\!=\!52$--$58$.
This constant trend is consistent with the weakly collective 
character of the normal configuration. In the lighter Sr 
isotopes, the low-lying $2^+$ state was proposed to contain 
a proton component associated with the quenching 
of the $\pi p_{1/2}$-$\pi p_{3/2}$ spacing as the $\nu d_{5/2}$ 
orbit is filled \cite{Federman1984}, in contrast to the 
predominantly neutron character of the corresponding 
low-lying $2^+$ states in the neighboring Zr isotopes 
\cite{Gavrielov2019, Gavrielov2020, Gavrielov2022}.
At neutron number 60 there is an abrupt drop from 
$0.815$~MeV (neutron number 58) to $0.145$~MeV, which 
marks a transition in the ground state configuration 
from the single-particle character to a collective-deformed 
one, which is now intruder, similarly to the transition in 
the zirconium chain.

The excited $0^+$ state descends steadily with neutron 
number, from about 2.7~MeV at 52 to 1.2~MeV 
at 58, where it is of intruder character. Then, at neutron 
number 60, it drops to 0.2~MeV and becomes normal. This 
abrupt change signals the Type~II QPT configuration 
crossing, and is a clear manifestation of the critical 
point between neutron number 58 ($^{96}$Sr) and 60 
($^{98}$Sr), similar to the zirconium isotopes $^{98}$Zr 
and $^{100}$Zr. The intruder band structure simultaneously 
evolves from a near-spherical, vibrational-like pattern in 
$^{90\text{--}96}$Sr to a compressed rotational-like 
spacing in $^{98,100}$Sr, signaling the Type~I QPT within 
it. At neutron number 62 ($^{100}$Sr), the $0^+_2$ rises 
again and may be interpreted as a $\beta$ bandhead state, 
similarly to $^{102}$Zr \cite{Gavrielov2019, Gavrielov2020, 
Gavrielov2022}.

\begin{figure}[tb!]
\centering
\includegraphics[width=1\linewidth]{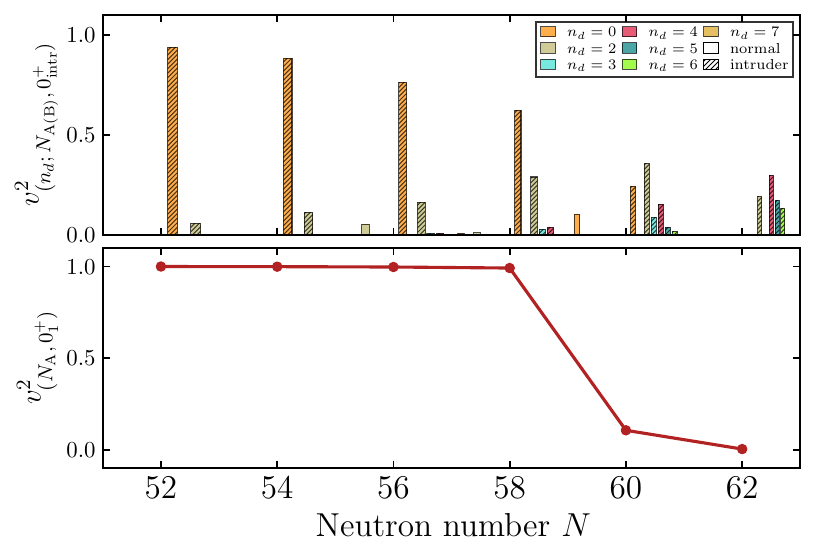}
\caption{
\textbf{(Upper panel)} Evolution of $n_d$ occupation 
probability of the lowest $0^+$ intruder state, 
\cref{eq:prob_nd}, depicting the different possible 
normal ($n_d=0\ldots,N$) and intruder 
($n_d=0\ldots,N+2$) $n_d$ values, which lie on the left and 
right of the neutron number, respectively, with intruder 
bars filled with diagonal lines. Each color represents a 
different $n_d$ number.
\textbf{(Lower panel)} Evolution of the amount of normal 
occupation in the wave function, \cref{eq:prob_bos}. 
Note the transition from normal in $^\text{90--96}$Sr to 
intruder in $^\text{98--100}$Sr.
\label{fig:decomp}}
\end{figure}

In order to identify both Type~I and II QPTs, we look at 
the structure of the resulting wave functions of the 
calculation. Figure~\ref{fig:decomp} shows 
the boson \eqref{eq:prob_bos} and $n_d$ \eqref{eq:prob_nd} 
decompositions for the ground state, $0^+_1$, and lowest 
intruder state, dubbed $0^+_\text{intr}$, respectively. In 
the figure, we see 
the normal content $v^2_{(N_A,0^+_1)}$ exceeds $99\%$ for 
$^{90\text{--}96}$Sr, then inverts abruptly at $^{98}$Sr, 
where the ground state is about $89\%$ intruder. By 
$^{100}$Sr it is essentially pure intruder.
The abrupt inversion in $v^2_{(N_A,0^+_1)}$ between 
$^{96}$Sr and $^{98}$Sr suggests the occurrence of a 
Type~II QPT --- the minimum of the potential-energy surface 
of one configuration crosses that of the other \cite{Frank2006}. 
The mixing between configurations remains small throughout, 
consistent with the weak configuration mixing observed 
experimentally in Refs.~\cite{Clement2016, Clement2016a} and 
in their two-state mixing analysis.

In order to unveil the Type~I QPT within the 
intruder band, it is insightful to examine the $n_d$ 
decomposition of the lowest $0^+$ intruder state, as shown 
in the upper panel of \cref{fig:decomp}. For 
$^\text{90--96}$Sr, the $0^+$ intruder state has a dominant 
$n_d\!=\!0$ component, reflecting a near-spherical 
structure. 
The strength of this component gradually descends, as those 
of other $n_d \not= 0$ components increase until, at 
$^{98}$Sr, we observe a strong mixing between several $n_d$ 
components, as occurs for a deformed state. This gradual 
change in the $n_d$ distribution, concurrent with the 
abrupt crossing in $v^2_{(N_A,0^+_1)}$, establishes the 
occurrence of IQPTs.

\begin{figure}[tb!]
\centering
\includegraphics[width=1\linewidth]{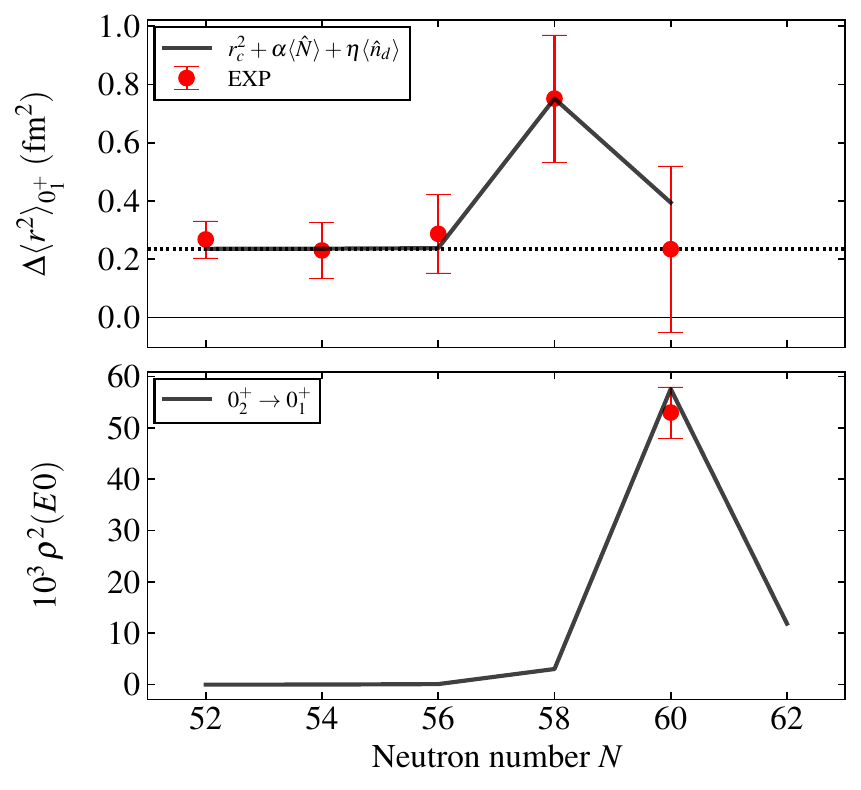}
\caption{Upper panel: isotope shift. Lower panel: monopole 
transitions. Data taken from \cite{Angeli2013, Kibedi2022}.
\label{fig:iso-e0}}
\end{figure}
Another notable observable that can help identify the 
abrupt structural changes in the strontium isotopes is the 
charge radius 
\begin{equation}\label{eq:charge-r}
\hat T(r^2) = r^2_c + \alpha \hat N_b + \eta \hat n_d~,
\end{equation}
where $r^2_c$ is the square radius of the closed shell, 
$\hat N_b$ ($\hat n_d$) is the total boson ($d$-boson) 
number 
operator~\cite{IachelloArimaBook}.
Using the charge radius \eqref{eq:charge-r}, one can 
calculate the isotope shift $\Delta\braket{\hat r^2}_{0^+_1}
=\braket{\hat{r}^{2}}_{0^{+}_1;A+2}-\braket{\hat{r}^2}_{0^{+}_1;A}$,
where $\braket{\hat r^2}_{0^+_1} $ is the expectation value 
of $\hat r^2$ in the ground state, $0^+_1$.
It depends on two parameters, $\alpha$ and $\eta$, given in 
units of fm$^2$. The parameter $\alpha$ represents the 
smooth behavior in $\Delta \braket{\hat{r}^{2}}_{0^{+}_1}$ 
due to the $A^{1/3}$ increase of the nuclear radius, while 
$\eta$ takes into account the effect of deformation. Their 
values are determined through the procedure in 
Ref.~\cite{Zerguine2012} and yield $\alpha\!=\!0.235$ 
fm$^2$, which takes the same value as for the adjacent 
zirconium isotopes due to the common spherical behavior of 
the ground state \cite{Gavrielov2019,Gavrielov2022}, and 
$\eta\!=\!0.053$ fm$^2$, which is about half the zirconium 
value (0.12 fm$^2$).
In the upper panel of \cref{fig:iso-e0}, the experimental 
and calculated $\Delta \braket{\hat r^2}_{0^+_1}$ values 
lie approximately on a straight line for $^\text{90--94}$Sr 
due to the constant value of $\alpha$ and lack of 
deformation, which gives approximately $\braket{\hat 
n_d}_{0^+_1} \approx 0$. 
At $^{98}$Sr we see a peak at the transition point, 
followed by a decrease in $^{100}$Sr, which is the expected 
behavior of a Type~II (first-order) QPT.
As we can see, the calculation reproduces the data along 
the entire chain and is within the error bars. 

%%%%%%%%%%%%%%%%%%%%%%%%%%%%%%%%%%%%%%%%%%%%%%%%%%%%%%%%%%%%%%%%%
\begin{figure*}[tb!]
\centering
\includegraphics[width=0.24\linewidth]{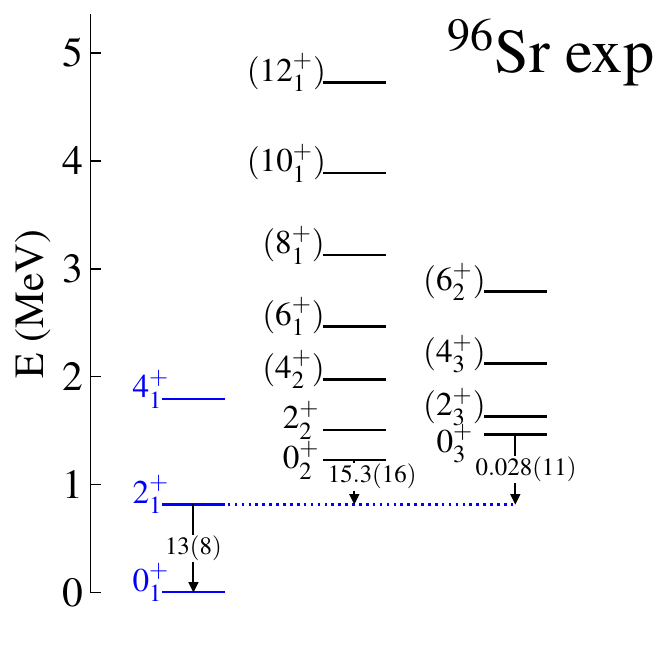}
\includegraphics[width=0.24\linewidth]{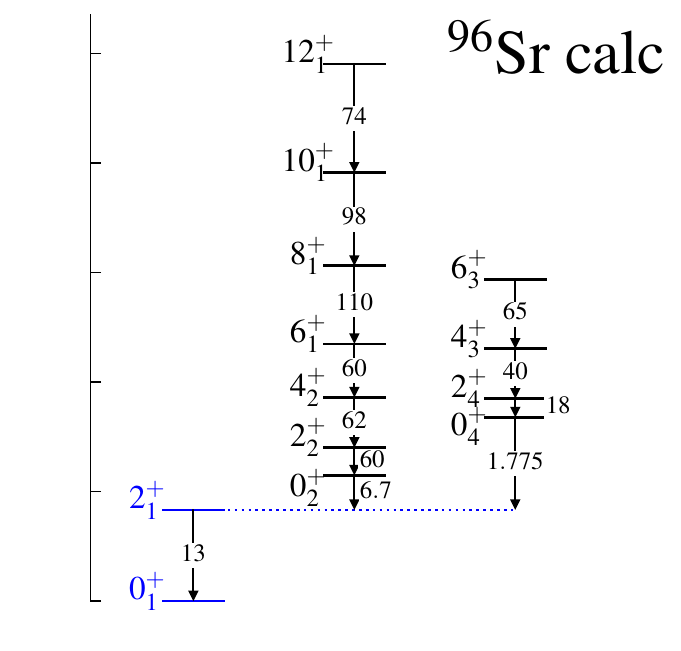}
\includegraphics[width=0.24\linewidth]{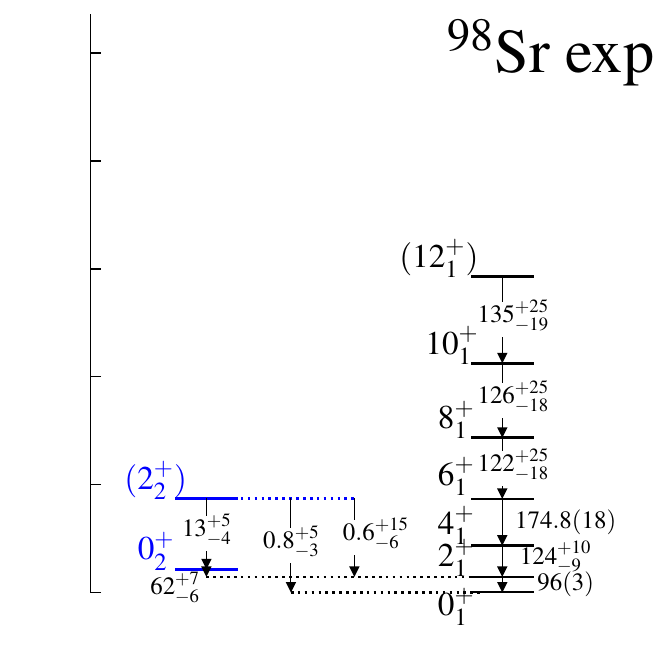}
\includegraphics[width=0.24\linewidth]{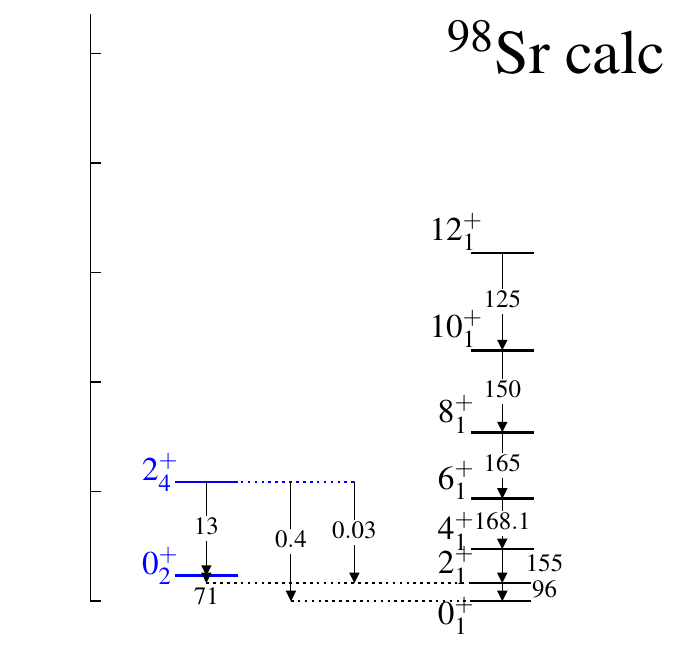}
\caption{Comparison between experimental (filled markers) 
and calculated (empty markers) lowest-energy levels in 
even-even Sr isotopes. Markers in red denote extrapolated 
experimental values and correspond to the state with the 
same marker.
\label{fig:96-98Sr}}
\end{figure*}
%%%%%%%%%%%%%%%%%%%%%%%%%%%%%%%%%%%%%%%%%%%%%%%%%%%%%%%%%%%%%%%%%

The charge-radius operator can also be used to evaluate the 
square of the monopole strength for $E0$ transitions 
between initial $\ket{i}$ and final $\ket{f}$ states,
\begin{equation}\label{eq:rho}
\rho(E0)=\frac{\braket{f|\hat T(E0)|i}}{eR^2}~.
\end{equation}
It can be evaluated using the $E0$ transition operator
\begin{equation}\label{eq:e0}
\hat T(E0)=(e_nN+e_pZ)\hat T(r^2).
\end{equation}
The latter is constructed from the charge radius operator 
\eqref{eq:charge-r}, in the manner suggested in 
\cite{Zerguine2008, Zerguine2012}.
Thus, the parameters used in \cref{eq:e0} take the same 
values as the parameters used for the isotope shift. For the 
strontium isotopes, shown in the lower panel of 
\cref{fig:iso-e0}, only one experimental 
data point is available, for the $0^+_2 \to 0^+_1$ 
transition, which the calculation predicts within the error 
bars. It shows a peak in the strength at $^{98}$Sr, 
denoting the abrupt change in structure and signaling the 
occurrence of a QPT. The peak arises due to the increased 
mixing between the normal and intruder $0^+$ states, after 
the crossing, where the difference in energy between them 
is minimal along the chain. The experimental value of 53(5) 
for the $0^+_2 \to 0^+_1$ transition of $^{98}$Sr is about 
half the value of 108(19) for $^{100}$Zr, which explains 
the roughly factor-of-two smaller value in $\eta$.

The critical point of the IQPT in the strontium isotopes is 
where the intruder configuration evolves in shape from 
spherical to deformed and simultaneously crosses the normal 
configuration, turning the ground state into an 
intruder-deformed state. This occurs between neutron 
numbers 58 ($^{96}$Sr) and 60 ($^{98}$Sr), and in order to 
obtain further insight into this region, we examine their 
individual spectra, shown in \Cref{fig:96-98Sr}. We can see 
the calculations reproduce the data to a good degree. For $^{96}$Sr, the 
$0^+_1$ and $2^+_1$ states of the normal configuration, 
shown in blue, are calculated to be almost purely normal, 
with $v^2_{(N_A)}>99\%$. The $4^+_1$ and $2^+_1$ states 
have an energy ratio of $E_{4^+_1}/E_{2^+_1} \approx 2.2$, 
which is typical for a spherical shape, and are assumed to 
be single-particle excitations \cite{Federman1984}. As such, 
the $4^+_1$ is assumed to lie outside the IBM-CM model 
space, as was similarly assumed for the zirconium isotopes 
\cite{Gavrielov2019,Gavrielov2020,Gavrielov2022}. The 
$0^+_2$ state is calculated to be a bandhead of an intruder 
band, which is weakly prolate-deformed, as the $n_d=0$ 
component (62\%) in \cref{fig:decomp} suggests. 
Furthermore, a dominant SO(6) component of 61\% is also
seen in the wave function (not shown in \cref{fig:decomp}), 
indicating further departure from the spherical U(5) limit. 
Alongside the lowest intruder $0^+_2$, the $0^+_4$ state seems to be 
an intruder $\beta$-bandhead, with the calculated 
$2^+_4,4^+_3,6^+_3$ states lying on top of it. The $n_d$ 
occupation of the $0^+_4$ state is spread among many 
components, without a single dominant one, suggesting a 
stronger deformation. However, it could also arise from an 
additional $4p$-$6h$ configuration, as was suggested in 
Ref.~\cite{Cruz2018}, which is outside the current 
model space.
While the mixing between the two configurations is 
calculated to be weak, the experimental $E2$ transition 
$0^+_2 \to 2^+_1$ is 15.3~W.u., which is not insignificant. 
The calculated value is 6.7~W.u., about half the 
experimental value. Taking a slightly stronger 
mixing parameter in the IBM-CM Hamiltonian \eqref{eq:W} 
reduces this discrepancy without changing the other 
transitions much. However, it is preferred to keep the mixing 
parameter $\omega$ constant along the chain, for 
simplicity. 
The intruder intraband transitions are calculated to be 
strong as expected from a deformed band and to decrease 
progressively with angular momentum, as expected in a 
finite boson model space \cite{IachelloArimaBook}.

For $^{98}$Sr, we see the $0^+_2$ state reaches its minimum 
energy. The large $B(E2;0_2^+ \to 2^+_1)$ value suggests 
the two configurations are more mixed and the crossing of 
the rotational band from $^{96}$Sr is apparent, i.e. a 
Type~II QPT has occurred. The yrast intruder band $0^+_1, 
2^+_1, 4^+_1, 6^+_1$ forms a strongly deformed rotational 
sequence with $E(4^+_1)/E(2^+_1) \!\approx\! 3.01$. The 
increasing $B(E2)$ values compared to $^{96}$Sr also 
suggest an increase in deformation, as the $n_d$-occupation 
of the $0^+_1$ wave function indicates in 
\cref{fig:decomp}. The $0^+_2$ state is suggested to 
belong to the normal configuration, with the calculated 
$2^+_4$ (corresponding to the experimental $(2^+_2)$ state) 
lying above it with an $E2$ transition that is comparable 
to the lighter strontium isotopes' normal $2^+ \to 0^+$ 
transition.
While the mixing between the configurations is calculated 
to be weak, with about $11\%$ normal component in the 
$0^+_1$ state, the $B(E2;0_2^+ \to 2^+_1) = 
62^{+7}_{-6}$~W.u. is surprisingly strong and reproduced by 
the calculation within the experimental uncertainty. This 
scenario is very similar to the $^{100}$Zr case 
\cite{Gavrielov2019, Gavrielov2022, Gavrielov2020}.

The spectrum of the even-even $^{90\text{--}100}$Sr 
isotopes was analyzed in the framework of the interacting 
boson model with configuration mixing. We examined the 
evolution of energy levels, the configuration and symmetry 
content of the wave functions, $E2$ transition rates, isotope 
shifts, and monopole $E0$ transition strengths. Special 
attention was given to the individual spectra of $^{96}$Sr 
and $^{98}$Sr. In general, the calculated results were found 
to reproduce the experimental data to a very good degree, by 
employing a simple approach for the Hamiltonian parameters, 
which kept most of them constant. 

The analysis reveals the occurrence of intertwined quantum 
phase transitions (IQPTs) in the strontium isotopes. It 
identifies a shape-evolution within the intruder 
configuration from a spherical to a deformed shape (Type~I 
QPT), and a crossing between the normal and intruder 
configurations (Type~II QPT) --- where the critical point 
of both QPTs lies between neutron number 58 ($^{96}$Sr) and 
60 ($^{98}$Sr), as in the even-even and odd-mass zirconium 
\cite{Gavrielov2019, Gavrielov2020, Gavrielov2022, 
Gavrielov2025} and the odd-mass niobium 
\cite{Gavrielov2022c, Gavrielov2023a} isotopes. 
The Type~II QPT occurs with weak mixing between the two 
configurations, keeping them almost pure before and after 
the crossing, which helps distinguish the two types of 
QPTs and identify the IQPT scenario.

This work adds another piece to the puzzle of the intricate 
$A \approx 100$ region, and thus reinforces a unified 
picture of intertwined quantum phase transitions (IQPTs). 
It motivates further studies in this and other regions to 
find similar occurrences of IQPTs. Specifically, 
it opens the way for analogous studies of the odd-mass 
isotopes of yttrium ($Z=39$) and strontium, by coupling a 
proton or neutron, respectively, to the strontium boson 
core. In such a scenario, the interplay between collective 
and single-particle degrees of freedom is expected to be 
even richer. Further experimental measurements would be 
valuable, especially of electromagnetic transition rates, 
which will help to clarify the collective structure, its 
deformation and configuration content. Special attention to 
$E2$ transitions from excited $2^+$ states and between the 
first intruder $2^+$ and $0^+$ states would be beneficial.

%%==========================================================
\nolinenumbers
\bibliography{refs.bib}

@article{Federman1977,
  author =        {Federman, P. and Pittel, S.},
  journal =       {Phys. Lett. B},
  number =        {4},
  pages =         {385--388},
  title =         {{Towards a unified microscopic description of nuclear
                   deformation}},
  volume =        {69},
  year =          {1977},
  doi =           {http://dx.doi.org/10.1016/0370-2693(77)90825-5},
  url =           {http://www.sciencedirect.com/science/article/pii/
                  0370269377908255},
}

@article{Federman1979,
  author =        {Federman, P. and Pittel, S.},
  journal =       {Phys. Rev. C},
  month =         {aug},
  number =        {2},
  pages =         {820--829},
  publisher =     {American Physical Society},
  title =         {{Unified shell-model description of nuclear
                   deformation}},
  volume =        {20},
  year =          {1979},
  doi =           {10.1103/PhysRevC.20.820},
  url =           {http://link.aps.org/doi/10.1103/PhysRevC.20.820},
}

@article{Leoni2024,
  author =        {Leoni, S. and Fornal, B. and Bracco, A. and
                   Tsunoda, Y. and Otsuka, Takaharu},
  journal =       {Prog. Part. Nucl. Phys.},
  month =         {may},
  pages =         {104119},
  publisher =     {Pergamon},
  title =         {{Multifaceted character of shape coexistence
                   phenomena in atomic nuclei}},
  year =          {2024},
  doi =           {10.1016/J.PPNP.2024.104119},
  issn =          {0146-6410},
  url =           {https://www.sciencedirect.com/science/article/pii/
                  S0146641024000231},
}

@article{Garrett2022,
  author =        {Garrett, Paul E. and Zieli{\'{n}}ska, Magda and
                   Cl{\'{e}}ment, Emmanuel},
  journal =       {Prog. Part. Nucl. Phys.},
  month =         {may},
  pages =         {103931},
  publisher =     {Pergamon},
  title =         {{An experimental view on shape coexistence in
                   nuclei}},
  volume =        {124},
  year =          {2022},
  doi =           {10.1016/J.PPNP.2021.103931},
  issn =          {0146-6410},
  url =           {https://www.sciencedirect.com/science/article/pii/
                  S0146641021000922?via%3Dihub},
}

@article{Cejnar2010,
  author =        {Cejnar, Pavel and Jolie, J. and Casten, R. F.},
  journal =       {Rev. Mod. Phys.},
  number =        {3},
  pages =         {2155--2212},
  publisher =     {American Physical Society},
  title =         {{Quantum phase transitions in the shapes of atomic
                   nuclei}},
  volume =        {82},
  year =          {2010},
  doi =           {10.1103/RevModPhys.82.2155},
  url =           {http://link.aps.org/doi/10.1103/RevModPhys.82.2155},
}

@article{Gilmore1978a,
  author =        {Gilmore, Robert and Feng, Da Hsuan},
  journal =       {Nucl. Phys. A},
  number =        {2},
  pages =         {189--204},
  title =         {{Phase transitions in nuclear matter described by
                   pseudospin Hamiltonians}},
  volume =        {301},
  year =          {1978},
  doi =           {http://dx.doi.org/10.1016/0375-9474(78)90260-9},
  url =           {http://www.sciencedirect.com/science/article/pii/
                  0375947478902609},
}

@article{Gilmore1979,
  author =        {Gilmore, Robert},
  journal =       {J. Math. Phys.},
  number =        {5},
  pages =         {891--893},
  title =         {{The classical limit of quantum nonspin systems}},
  volume =        {20},
  year =          {1979},
  doi =           {http://dx.doi.org/10.1063/1.524137},
  url =           {http://scitation.aip.org/content/aip/journal/jmp/20/5/
                  10.1063/1.524137},
}

@article{Frank2006,
  author =        {Frank, Alejandro and {Van Isacker}, P. and
                   Iachello, F.},
  journal =       {Phys. Rev. C},
  month =         {jun},
  number =        {6},
  pages =         {061302(R)},
  publisher =     {American Physical Society},
  title =         {{Phase transitions in configuration mixed models}},
  volume =        {73},
  year =          {2006},
  doi =           {10.1103/PhysRevC.73.061302},
  issn =          {0556-2813},
  url =           {https://link.aps.org/doi/10.1103/PhysRevC.73.061302},
}

@article{Heyde2011,
  author =        {Heyde, K. and Wood, John L.},
  journal =       {Rev. Mod. Phys.},
  month =         {nov},
  number =        {4},
  pages =         {1467--1521},
  publisher =     {American Physical Society},
  title =         {{Shape coexistence in atomic nuclei}},
  volume =        {83},
  year =          {2011},
  doi =           {10.1103/RevModPhys.83.1467},
  url =           {http://link.aps.org/doi/10.1103/RevModPhys.83.1467},
}

@article{Gavrielov2019,
  author =        {Gavrielov, N. and Leviatan, A. and Iachello, F.},
  journal =       {Phys. Rev. C},
  month =         {jun},
  number =        {6},
  pages =         {064324},
  publisher =     {American Physical Society},
  title =         {{Intertwined quantum phase transitions in the Zr
                   isotopes}},
  volume =        {99},
  year =          {2019},
  doi =           {10.1103/PhysRevC.99.064324},
  issn =          {2469-9985},
  url =           {https://link.aps.org/doi/10.1103/PhysRevC.99.064324},
}

@article{Gavrielov2020,
  author =        {Gavrielov, N. and Leviatan, A. and Iachello, F.},
  journal =       {Phys. Scr.},
  month =         {feb},
  number =        {2},
  pages =         {024001},
  publisher =     {IOP Publishing},
  title =         {{Interplay between shape-phase transitions and shape
                   coexistence in the Zr isotopes}},
  volume =        {95},
  year =          {2020},
  doi =           {10.1088/1402-4896/ab456b},
  issn =          {0031-8949},
  url =           {https://iopscience.iop.org/article/10.1088/1402-4896/
                  ab456b},
}

@article{Gavrielov2022,
  author =        {Gavrielov, N. and Leviatan, A. and Iachello, F.},
  journal =       {Phys. Rev. C},
  month =         {jan},
  number =        {1},
  pages =         {014305},
  publisher =     {American Physical Society},
  title =         {{Zr isotopes as a region of intertwined quantum phase
                   transitions}},
  volume =        {105},
  year =          {2022},
  doi =           {10.1103/PhysRevC.105.014305},
  issn =          {2469-9985},
  url =           {https://link.aps.org/doi/10.1103/PhysRevC.105.014305},
}

@article{Gavrielov2022c,
  author =        {Gavrielov, N. and Leviatan, A. and Iachello, F.},
  journal =       {Phys. Rev. C},
  month =         {nov},
  number =        {5},
  pages =         {L051304},
  publisher =     {American Physical Society},
  title =         {{Mixed configurations and intertwined quantum phase
                   transitions in odd-mass nuclei}},
  volume =        {106},
  year =          {2022},
  doi =           {10.1103/PhysRevC.106.L051304},
  issn =          {24699993},
  url =           {https://journals.aps.org/prc/abstract/10.1103/
                  PhysRevC.106.L051304},
}

@article{Gavrielov2023a,
  author =        {Gavrielov, N.},
  journal =       {Phys. Rev. C},
  month =         {jul},
  number =        {1},
  pages =         {014320},
  publisher =     {American Physical Society},
  title =         {{Configuration mixing and intertwined quantum phase
                   transitions in odd-mass niobium isotopes}},
  volume =        {108},
  year =          {2023},
  doi =           {10.1103/PhysRevC.108.014320},
  issn =          {24699993},
  url =           {https://link.aps.org/doi/10.1103/PhysRevC.108.014320},
}

@article{Gavrielov2025,
  author =        {Gavrielov, N.},
  journal =       {Phys. Rev. Res.},
  month =         {apr},
  number =        {2},
  pages =         {L022022},
  publisher =     {American Physical Society},
  title =         {{Competing shape evolution, crossing configurations,
                   and single particle levels in nuclei}},
  volume =        {7},
  year =          {2025},
  doi =           {10.1103/PhysRevResearch.7.L022022},
  issn =          {26431564},
  url =           {https://link.aps.org/doi/10.1103/PhysRevResearch.7.L022022},
}

@article{Mach1991,
  author =        {Mach, H. and Wohn, F. K. and Moln{\'{a}}r, G. and
                   Sistemich, K. and Hill, John C. and
                   Moszy{\'{n}}ski, M. and Gill, R. L. and Krips, W. and
                   Brenner, D. S.},
  journal =       {Nucl. Phys. A},
  month =         {feb},
  number =        {2},
  pages =         {197--227},
  publisher =     {North-Holland},
  title =         {{Retardation of B(E2; 01+ → 21+) rates in 90–96Sr
                   and strong subshell closure effects in the A $\sim$
                   100 region}},
  volume =        {523},
  year =          {1991},
  doi =           {10.1016/0375-9474(91)90001-M},
  issn =          {0375-9474},
  url =           {https://www.sciencedirect.com/science/article/pii/
                  037594749190001M},
}

@article{Clement2016,
  author =        {Cl{\'{e}}ment, Emmanuel and Zieli\`{n}ska, M. and
                   G{\"{o}}rgen, A. and Korten, W. and P{\'{e}}ru, S. and
                   Libert, J. and Goutte, H. and Hilaire, S. and
                   Bastin, B. and Bauer, C. and Blazhev, A. and Bree, N. and
                   Bruyneel, B. and Butler, P. A. and Butterworth, J. and
                   Delahaye, P. and Dijon, A. and Doherty, D. T. and
                   Ekstr{\"{o}}m, A. and Fitzpatrick, C. and Fransen, C. and
                   Georgiev, G. and Gernh{\"{a}}user, R. and Hess, H. and
                   Iwanicki, J. and Jenkins, David Gareth and
                   Larsen, A. C. and Ljungvall, J. and Lutter, R. and
                   Marley, P. and Moschner, K. and Napiorkowski, P. J. and
                   Pakarinen, J. and Petts, A. and Reiter, P. and
                   Renstr{\o}m, T. and Seidlitz, M. and Siebeck, B. and
                   Siem, S. and Sotty, C. and Srebrny, J. and
                   Stefanescu, I. and Tveten, G. M. and
                   {Van de Walle}, J. and Vermeulen, M. and Voulot, D. and
                   Warr, N. and Wenander, F. and Wiens, A. and
                   {De Witte}, H. and Wrzosek-Lipska, K.},
  journal =       {Phys. Rev. Lett.},
  month =         {jan},
  number =        {2},
  pages =         {022701},
  publisher =     {American Physical Society},
  title =         {{Spectroscopic Quadrupole Moments in ${96,98}^Sr$:
                   Evidence for Shape Coexistence in Neutron-Rich
                   Strontium Isotopes at $N = 60$}},
  volume =        {116},
  year =          {2016},
  doi =           {10.1103/PhysRevLett.116.022701},
  url =           {http://link.aps.org/doi/10.1103/PhysRevLett.116.022701},
}

@article{Clement2016a,
  author =        {Cl{\'{e}}ment, Emmanuel and Zieli\`{n}ska, M. and
                   P{\'{e}}ru, S. and Goutte, H. and Hilaire, S. and
                   G{\"{o}}rgen, A. and Korten, W. and Doherty, D. T. and
                   Bastin, B. and Bauer, C. and Blazhev, A. and Bree, N. and
                   Bruyneel, B. and Butler, P. A. and Butterworth, J. and
                   Cederk{\"{a}}ll, J. and Delahaye, P. and Dijon, A. and
                   Ekstr{\"{o}}m, A. and Fitzpatrick, C. and Fransen, C. and
                   Georgiev, G. and Gernh{\"{a}}user, R. and Hess, H. and
                   Iwanicki, J. and Jenkins, David Gareth and
                   Larsen, A. C. and Ljungvall, J. and Lutter, R. and
                   Marley, P. and Moschner, K. and Napiorkowski, P. J. and
                   Pakarinen, J. and Petts, A. and Reiter, P. and
                   Renstr{\o}m, T. and Seidlitz, M. and Siebeck, B. and
                   Siem, S. and Sotty, C. and Srebrny, J. and
                   Stefanescu, I. and Tveten, G. M. and
                   {Van de Walle}, J. and Vermeulen, M. and Voulot, D. and
                   Warr, N. and Wenander, F. and Wiens, A. and
                   {De Witte}, H. and Wrzosek-Lipska, K.},
  journal =       {Phys. Rev. C},
  month =         {nov},
  number =        {5},
  pages =         {054326},
  publisher =     {American Physical Society},
  title =         {{Low-energy Coulomb excitation of $^{96 , 98}$Sr
                   beams}},
  volume =        {94},
  year =          {2016},
  doi =           {10.1103/PhysRevC.94.054326},
  issn =          {2469-9985},
  url =           {https://link.aps.org/doi/10.1103/PhysRevC.94.054326},
}

@article{MayaBarbecho2022,
  author =        {Maya-Barbecho, E. and Garc\'ia-Ramos, J. E.},
  journal =       {Phys. Rev. C},
  month =         {mar},
  number =        {3},
  pages =         {034341},
  publisher =     {American Physical Society},
  title =         {{Shape coexistence in Sr isotopes}},
  volume =        {105},
  year =          {2022},
  doi =           {10.1103/PhysRevC.105.034341},
  issn =          {2469-9985},
  url =           {https://link.aps.org/doi/10.1103/PhysRevC.105.034341},
}

@article{Nomura2016c,
  author =        {Nomura, Kosuke and Rodriguez-Guzman, R. and
                   Robledo, Luis M.},
  journal =       {Phys. Rev. C},
  month =         {oct},
  number =        {4},
  pages =         {044314},
  publisher =     {American Physical Society},
  title =         {{Structural evolution in $A\approx100$ nuclei within
                   the mapped interacting boson model based on the Gogny
                   energy density functional}},
  volume =        {94},
  year =          {2016},
  doi =           {10.1103/PhysRevC.94.044314},
  url =           {https://link.aps.org/doi/10.1103/PhysRevC.94.044314},
}

@article{Regis2017c,
  author =        {R\'egis, J.-M. and Jolie, J. and Saed-Samii, N. and
                   Warr, N. and Pfeiffer, M. and Blanc, A. and
                   Jentschel, M. and K{\"{o}}ster, U. and Mutti, P. and
                   Soldner, T. and Simpson, G. S. and Drouet, F. and
                   Vancraeyenest, A. and de France, G. and
                   Cl{\'{e}}ment, Emmanuel and Stezowski, O. and
                   Ur, C. A. and Urban, W. and Regan, P. H. and
                   Podoly\`{a}k, Zs. and Larijani, C. and Townsley, C. and
                   Carroll, R. and Wilson, E. and Fraile, L. M. and
                   Mach, H. and Paziy, V. and Olaizola, B. and Vedia, V. and
                   Bruce, A. M. and Roberts, O. J. and Smith, J. F. and
                   Scheck, M. and Kr{\"{o}}ll, T. and Hartig, A.-L. and
                   Ignatov, A. and Ilieva, S. and Lalkovski, S. and
                   Korten, W. and M\u{a}rginean, N. and Otsuka, Takaharu and
                   Shimizu, Noritaka and Togashi, Tomoaki and
                   Tsunoda, Yusuke},
  journal =       {Phys. Rev. C},
  month =         {may},
  number =        {5},
  pages =         {054319},
  publisher =     {American Physical Society},
  title =         {{Abrupt shape transition at neutron number $N=60$:
                   $B(E2)$ values in $^{94,96,98}$Sr from fast
                   $\gamma-gamma$ timing}},
  volume =        {95},
  year =          {2017},
  doi =           {10.1103/PhysRevC.95.054319},
  url =           {http://link.aps.org/doi/10.1103/PhysRevC.95.054319},
}

@article{Duval1981,
  author =        {Duval, Philip D. and Barrett, Bruce R.},
  journal =       {Phys. Lett. B},
  number =        {3},
  pages =         {223--227},
  title =         {{Configuration mixing in the interacting boson
                   model}},
  volume =        {100},
  year =          {1981},
  doi =           {10.1016/0370-2693(81)90321-X},
}

@article{Duval1982,
  author =        {Duval, Philip D. and Barrett, Bruce R.},
  journal =       {Nucl. Phys. A},
  number =        {2},
  pages =         {213--228},
  publisher =     {North-Holland},
  title =         {{Quantitative description of configuration mixing in
                   the interacting boson model}},
  volume =        {376},
  year =          {1982},
  doi =           {10.1016/0375-9474(82)90061-6},
  url =           {http://www.sciencedirect.com/science/article/pii/
                  0375947482900616},
}

@article{Heyde1985,
  author =        {Heyde, K. and {Van Isacker}, P. and Casten, R. F. and
                   Wood, John L.},
  journal =       {Phys. Lett. B},
  number =        {5--6},
  pages =         {303--308},
  title =         {{A shell-model interpretation of intruder states and
                   the onset of deformation in even-even nuclei}},
  volume =        {155},
  year =          {1985},
  doi =           {http://dx.doi.org/10.1016/0370-2693(85)91575-8},
  issn =          {0370-2693},
  url =           {http://www.sciencedirect.com/science/article/pii/
                  0370269385915758},
}

@article{Heyde1987,
  author =        {Heyde, K. and Jolie, J. and Moreau, J and
                   Ryckebusch, J and Waroquier, M. and Duppen, P Van and
                   Huyse, M. and Wood, John L.},
  journal =       {Nucl. Phys. A},
  number =        {2},
  pages =         {189--226},
  title =         {{A shell-model description of 0+ intruder states in
                   even-even nuclei}},
  volume =        {466},
  year =          {1987},
  doi =           {http://dx.doi.org/10.1016/0375-9474(87)90439-8},
  issn =          {0375-9474},
  url =           {http://www.sciencedirect.com/science/article/pii/
                  0375947487904398},
}

@article{Ginocchio1980a,
  author =        {Ginocchio, Joseph N. and Kirson, Michael W.},
  journal =       {Phys. Rev. Lett.},
  number =        {26},
  pages =         {1744--1747},
  publisher =     {American Physical Society},
  title =         {{Relationship between the Bohr Collective Hamiltonian
                   and the Interacting-Boson Model}},
  volume =        {44},
  year =          {1980},
  doi =           {10.1103/PhysRevLett.44.1744},
  url =           {http://link.aps.org/doi/10.1103/PhysRevLett.44.1744},
}

@article{Ginocchio1980b,
  author =        {Ginocchio, Joseph N. and Kirson, Michael W.},
  journal =       {Nucl. Phys. A},
  number =        {1--2},
  pages =         {31--60},
  title =         {{An intrinsic state for the interacting boson model
                   and its relationship to the Bohr-Mottelson model}},
  volume =        {350},
  year =          {1980},
  doi =           {http://dx.doi.org/10.1016/0375-9474(80)90387-5},
  issn =          {0375-9474},
  url =           {http://www.sciencedirect.com/science/article/pii/
                  0375947480903875},
}

@article{Dieperink1980,
  author =        {Dieperink, A. E. L. and Scholten, Olaf and
                   Iachello, F.},
  journal =       {Phys. Rev. Lett.},
  number =        {26},
  pages =         {1747--1750},
  publisher =     {American Physical Society},
  title =         {{Classical Limit of the Interacting-Boson Model}},
  volume =        {44},
  year =          {1980},
  doi =           {10.1103/PhysRevLett.44.1747},
  url =           {http://link.aps.org/doi/10.1103/PhysRevLett.44.1747},
}

@misc{ensdf,
  author =        {{Evaluated Nuclear Structure Data File (ENSDF)}},
  booktitle =     {Eval. Nucl. Struct. Data File Search Retr.},
  title =         {https://www.nndc.bnl.gov/ensdf},
  url =           {https://www.nndc.bnl.gov/ensdf/},
}

@article{Cruz2020,
  author =        {Cruz, S. and Wimmer, K. and Bhattacharjee, S. S. and
                   Bender, P. C. and Hackman, G. and Kr{\"{u}}cken, R. and
                   Ames, F. and Andreoiu, C. and Austin, R. A.E. and
                   Bancroft, C. S. and Braid, R. and Bruhn, T. and
                   Catford, W. N. and Cheeseman, A. and Chester, A. and
                   Cross, D. S. and Diget, C. Aa and Drake, T. and
                   Garnsworthy, A. B. and Kanungo, R. and Knapton, A. and
                   Korten, W. and Kuhn, K. and Lassen, J. and Laxdal, R. and
                   Marchetto, M. and Matta, A. and Miller, D. and
                   Moukaddam, M. and Orr, N. A. and Sachmpazidi, N. and
                   Sanetullaev, A. and Svensson, C. E. and Terpstra, N. and
                   Unsworth, C. and Voss, P. J.},
  journal =       {Phys. Rev. C},
  month =         {aug},
  number =        {2},
  pages =         {024335},
  publisher =     {American Physical Society},
  title =         {{Single-particle structure in neutron-rich Sr
                   isotopes approaching $N=60$ the shape transition}},
  volume =        {102},
  year =          {2020},
  doi =           {10.1103/PhysRevC.102.024335},
  issn =          {24699993},
  url =           {https://journals.aps.org/prc/abstract/10.1103/
                  PhysRevC.102.024335},
}

@article{Marchini2026,
author = {Marchini, N. and Rocchini, M. and Zieli{\'{n}}ska, M. and Nannini, A. and Doherty, D.T. and Gavrielov, N. and Garrett, P.E. and Hady{\'{n}}ska-Kl{\k{e}}k, K. and Goasduff, A. and Testov, D. and Bakes, S.D. and Bazzacco, D. and Benzoni, G. and Berry, T. and Brugnara, D. and Camera, F. and Catford, W.N. and Chiari, M. and Galtarossa, F. and Gelli, N. and Gottardo, A. and Gozzelino, A. and Illana, A. and Keatings, J. and Mengoni, D. and Morrison, L. and Napoli, D.R. and Ottanelli, M. and Ottanelli, P. and Pasqualato, G. and Recchia, F. and Riccetto, S. and Scheck, M. and Siciliano, M. and {Valiente D{\'{o}}bon}, J.J. and Zanon, I.},
doi = {10.1016/j.physletb.2026.140668},
issn = {03702693},
journal = {Phys. Lett. B},
month = {aug},
pages = {140668},
publisher = {North-Holland},
title = {{Shape coexistence in 94Zr from a model-independent analysis}},
url = {https://www.sciencedirect.com/science/article/pii/S0370269326005204},
volume = {879},
year = {2026}
}

@article{Federman1984,
  author =        {Federman, P. and Pittel, S. and Etchegoyen, A.},
  journal =       {Phys. Lett. B},
  month =         {jun},
  number =        {5-6},
  pages =         {269--271},
  publisher =     {North-Holland},
  title =         {{Quenching of the $2p_{1/2}-2p_{3/2}$ proton
                   spin-orbit splitting in the Sr-Zr region}},
  volume =        {140},
  year =          {1984},
  doi =           {10.1016/0370-2693(84)90750-0},
  issn =          {0370-2693},
  url =           {https://www.sciencedirect.com/science/article/pii/
                  0370269384907500},
}

@article{Severyukhin2018,
  author =        {Severyukhin, A. P. and Arsenyev, N. N. and
                   Pietralla, N. and Werner, V.},
  journal =       {Eur. Phys. J. A},
  month =         {jan},
  number =        {1},
  pages =         {4},
  publisher =     {Springer Berlin Heidelberg},
  title =         {{Proton-neutron structure of first and second
                   quadrupole excitations of 90Sr}},
  volume =        {54},
  year =          {2018},
  doi =           {10.1140/epja/i2018-12451-4},
  issn =          {1434-6001},
  url =           {http://link.springer.com/10.1140/epja/i2018-12451-4},
}

@article{Angeli2013,
  author =        {Angeli, I. and Marinova, Krassimira p.},
  journal =       {At. Data Nucl. Data Tables},
  month =         {jan},
  number =        {1},
  pages =         {69--95},
  publisher =     {Academic Press},
  title =         {{Table of experimental nuclear ground state charge
                   radii: An update}},
  volume =        {99},
  year =          {2013},
  doi =           {10.1016/j.adt.2011.12.006},
  isbn =          {0092-640X},
  issn =          {0092640X},
  url =           {http://www.sciencedirect.com/science/article/pii/
                  S0092640X12000265},
}

@article{Kibedi2022,
  author =        {Kib{\'{e}}di, T. and Garnsworthy, A. B. and
                   Wood, John L.},
  journal =       {Prog. Part. Nucl. Phys.},
  month =         {mar},
  pages =         {103930},
  publisher =     {Pergamon},
  title =         {{Electric monopole transitions in nuclei}},
  volume =        {123},
  year =          {2022},
  doi =           {10.1016/J.PPNP.2021.103930},
  issn =          {0146-6410},
  url =           {https://www.sciencedirect.com/science/article/pii/
                  S0146641021000910},
}

@book{IachelloArimaBook,
  author =        {Iachello, F. and Arima, Akito},
  publisher =     {Cambridge University Press},
  title =         {{The Interacting Boson Model}},
  year =          {1987},
  isbn =          {9780521302821},
  url =           {http://www.cambridge.org/il/academic/subjects/physics/
                  theoretical-physics-and-mathematical-physics/interacting-
                  boson-model?format=HB&isbn=9780521302821},
}

@article{Zerguine2012,
  author =        {Zerguine, S. and {Van Isacker}, P. and
                   Bouldjedri, A.},
  journal =       {Phys. Rev. C},
  month =         {mar},
  number =        {3},
  pages =         {034331},
  publisher =     {American Physical Society},
  title =         {{Consistent description of nuclear charge radii and
                   electric monopole transitions}},
  volume =        {85},
  year =          {2012},
  doi =           {10.1103/PhysRevC.85.034331},
  url =           {https://link.aps.org/doi/10.1103/PhysRevC.85.034331},
}

@article{Zerguine2008,
  author =        {Zerguine, S. and {Van Isacker}, P. and Bouldjedri, A. and
                   Heinze, S.},
  journal =       {Phys. Rev. Lett.},
  month =         {jul},
  number =        {2},
  pages =         {022502},
  publisher =     {American Physical Society},
  title =         {{Correlating Radii and Electric Monopole Transitions
                   of Atomic Nuclei}},
  volume =        {101},
  year =          {2008},
  doi =           {10.1103/PhysRevLett.101.022502},
  url =           {https://link.aps.org/doi/10.1103/PhysRevLett.101.022502},
}

@article{Cruz2018,
  author =        {Cruz, S. and Bender, P.C. and Kr{\"{u}}cken, R. and
                   Wimmer, K. and Ames, F. and Andreoiu, C. and
                   Austin, R.A.E. and Bancroft, C.S. and Braid, R. and
                   Bruhn, T. and Catford, W.N. and Cheeseman, A. and
                   Chester, A. and Cross, D.S. and Diget, C.Aa. and
                   Drake, T. and Garnsworthy, A.B. and Hackman, G. and
                   Kanungo, R. and Knapton, A. and Korten, W. and
                   Kuhn, K. and Lassen, J. and Laxdal, R. and
                   Marchetto, M. and Matta, A. and Miller, D. and
                   Moukaddam, M. and Orr, N.A. and Sachmpazidi, N. and
                   Sanetullaev, A. and Svensson, C.E. and Terpstra, N. and
                   Unsworth, C. and Voss, P.J.},
  journal =       {Phys. Lett. B},
  month =         {nov},
  pages =         {94--99},
  publisher =     {North-Holland},
  title =         {{Shape coexistence and mixing of low-lying 0+ states
                   in 96Sr}},
  volume =        {786},
  year =          {2018},
  doi =           {10.1016/J.PHYSLETB.2018.09.031},
  issn =          {0370-2693},
  url =           {https://www.sciencedirect.com/science/article/pii/
                  S0370269318307329?via%3Dihub},
}
%%==========================================================

\end{document}